\documentclass[aps,prx,preprint,groupedaddress]{revtex4-2}

\usepackage{listings}
\usepackage{amssymb}
\usepackage{amsmath}
\usepackage{physics}
\usepackage{mathtools}
\usepackage{empheq}
\usepackage{algorithm} 
\usepackage{algpseudocode}  
\usepackage[T1]{fontenc}

\usepackage{hyperref}
\usepackage{xcolor}

\newcommand{\br}[1]{\left(#1\right)}

\newcommand{\RNum}[1]{\uppercase\expandafter{\romannumeral #1\relax}}

\begin{document}

\title{Inverse Design of Tightly Woven Smart Fabrics}

\author{Einav Berin}
\author{Hillel Aharoni}
\email[]{hillel.aharoni@weizmann.ac.il}

\affiliation{Department of Complex Systems, Weizmann Institute of Science, Rehovot 7610001, Israel}

\date{\today}

\begin{abstract}
We present a geometric framework for the inverse design of smart woven fabrics composed of non-uniformly shrinking threads. A sufficiently tight weaving structure imposes strong local criteria on the material deformation and reduces the local geometry to a single scalar degree of freedom. Control over this degree of freedom can be achieved through simple calibration for each specific material system, via either mechanical experiments or numerical simulations. This reduction allows us to inverse-design a woven smart fabric, that conforms to an arbitrary target geometry when actuated, by solving a nonlinear hyperbolic partial differential equation. We validate this approach by deriving the thread-level actuation required for specific target geometries. We present both exact analytic solutions for symmetric shapes and a numerical optimization method for arbitrary freeform surfaces. Our manuscript provides motivation for weaving smart threads into fabrics, and provides a practical framework for achieving programmable, complex three-dimensional shaping of these novel metamaterials.
\end{abstract}

\maketitle

\section*{Introduction}

Fabrics are everywhere. They are lightweight, durable, scalable, and versatile materials, used in a wide variety of technologies, applications, and everyday uses since ancient times. They are composed of yarns or threads interlaced into a two-dimensional periodic structure. The threads are held together by the topological constraints of the interlacing, which depend on the specific interlacing process, such as weaving, knitting, crocheting, and others.
Several strategies have been developed for modeling fabrics, including thread-path tracking \cite{zheng2015overview}, mass–spring and finite-element models \cite{Provot1995, BaraffWitkin1998}, and continuum formulations that approximate the fabric as a two-dimensional surface \cite{Womersley01031937,Mokhtar}.
Some fabrics are made loose, attributing them with flexibility and the ability to stretch and shear. Others are made tight, typically by introducing tension during manufacturing, endowing them with a more robust geometry \cite{Boisse2017BiasReview}.

Extensive efforts have been invested in recent decades to develop smart materials that undergo controlled deformations in response to external stimuli \cite{Kim2020Shape}. Examples include hydrogels \cite{hirokawa1984volume}, nematic elastomers \cite{kupfer1991nematic}, shape-memory polymers \cite{lendlein2001ab}, and more. 
Recent advances in 3D printing have enabled the fabrication of smart threads that can be actuated by temperature \cite{kotikian20183d}, light, or electric fields \cite{liuShaping2020}. Recently, threads with a spatially varying actuation profile were demonstrated \cite{TellesLewis}. Using smart threads enables the manufacturing of woven, knitted, and embroidered smart fabrics \cite{Terentjev, Sun, Shu}.

Programmable surfaces made of smart materials are used in soft robotics \cite{Multigaitsoftrobo8}, biomedicine \cite{CrawfordGregoryP2007Lmfa}, 4D printing \cite{4d}, and the fashion industry \cite{KonakoviMina2016Bdcd}.
A central challenge in studying such surfaces is understanding the relationship between local actuation and global shape. This can be framed as two complementary problems. The forward problem asks, given the local material response at every point, what global shape will a surface assume upon actuation? The inverse problem seeks the local material response that needs to be programmed into the material at every point to achieve a desired target global shape. These problems have been studied in various systems, including two-dimensional liquid crystal elastomer sheets \cite{Aharoni7206}, auxetic materials \cite{KonakoviMina2016Bdcd}, responsive gel sheets \cite{KimJungwook2012DRBS}, deployable origami \cite{DANG2022111224}, and others.
Another structure often used to realize curved surfaces is that of triaxial weaves \cite{baek2021smooth, 3Dweavingwithcurvedribbons}. However, these structures are typically rigid (hyperstatic) and depend on curvature that arises from topological defects permanently embedded in the weaving pattern. As a result, they are less suited for shape-morphing or self-deployment applications compared to initially compliant biaxial fabrics.

\begin{figure}[h]
	\includegraphics[width=1.0\linewidth]{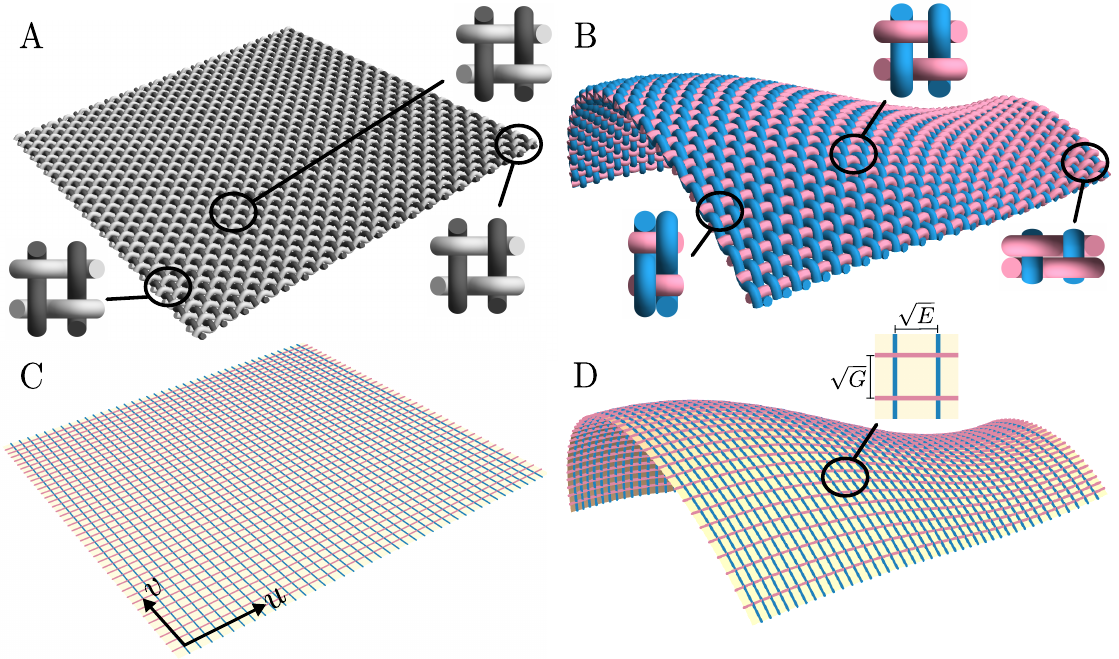}
	\caption{Illustration of a plain woven smart fabric. (A) A pre-actuated smart fabric. The fabric is flat and uniform (as shown by the identical zoomed-in unit cells).
		(B) When actuated, the spatially varying pre-programmed thread response deforms each unit cell differently, endowing the fabric with a non-Euclidean geometry that invokes its morphing into a non-planar surface.
		(C) Material coordinates $(u,v)$ aligned with warp and weft threads, respectively, are used to keep track of the fabric's geometry and shape.
		(D) The local tangent spacings of adjacent warp/weft threads define the metric scale factors $\sqrt{E}$ and $\sqrt{G}$. In a tightly woven fabric, shear is suppressed, and threads remain orthogonal.}   
	\label{fig: illustration of smart fabric}
\end{figure}

In this work, we focus on a system of a tightly woven smart fabric composed of threads with non-uniform shrinkage profiles, as shown in Fig. \ref{fig: illustration of smart fabric}. We focus on geometric constraints and do not consider dynamics or friction.
We demonstrate that the tight weave structure confines the range of local deformations to a single curve, which can be calibrated according to the working material. Constructing an orthogonal coordinate system based on the tight-weaving structure reveals a simple relation between the remaining controllable local degree of freedom and the actuated surface geometry. We utilize this approach to address both the forward and inverse design problems, producing explicit recipes for creating smart fabrics that deform into arbitrary desired shapes.

\section*{Results}

A plain woven fabric is a textile structure formed by the interlacing of warp and weft threads, where each weft thread passes over and under successive warp threads in an alternating sequence. We describe the configuration of a large-scale woven fabric as a two-dimensional (2D) surface representing the mid-plane of the structure, coarse-grained at the scale of the weaving unit cell. We use coordinates $u$ and $v$ that follow the woven threads; namely, $u$ counts the warp threads and $v$ the weft threads, measured from some arbitrary origin. We assume that the weaving unit-cell is much smaller than system-size or spatial gradients in the composition of the smart fabric, and thus describe the shape of the fabric in terms of a continuous function ${\bf f}\br{u,v}\in{\mathbb R}^3$. The two-dimensional geometry of the fabric at a given state is described by the metric tensor \cite{doCarmoManfredoP2016DGoC}
\begin{equation}
	\dd s^2 = E\br{u,v}\dd u^2 + 2F\br{u,v}\dd u \dd v + G\br{u,v}\dd v^2,
	\label{eq: def-metric}
\end{equation}
which captures the lateral dimensions of each unit cell in the weave structure, which we choose to measure in units of the thread's diameter (the lateral dimensions are typically a few times the thread diameter, thus in our units $E$ and $G$ are expected to exceed unity).

In the general case, the metric components $E,F,G$ may vary across the fabric and change when the fabric is deformed (either by external forcing or by actuation of the smart threads). In a tightly-woven fabric, however, threads lose their local freedom to change position or angle, and the fabric becomes rigid and stiff. Experiments show that increasing thread density rigidifies both bending and shear \cite{Roleoffibre,StructureandMechanicsofWovenFabrics} and diminishes the maximal shear angle \cite{Skelton}. Since the symmetry of the weaving unit-cell favors neither shear direction, and because we assume that lateral gradients at the scale of the unit cell are small, $F=0$ is imposed.
Recent variable-reed experiments further confirm that fabrics that are woven non-orthogonally under tension resume orthogonality (namely, $F=0$) when contracting to a (generally curved) jammed state \cite{cnaani2026weaving}.

\begin{figure}[h!]
	\includegraphics[width=1.0\linewidth]{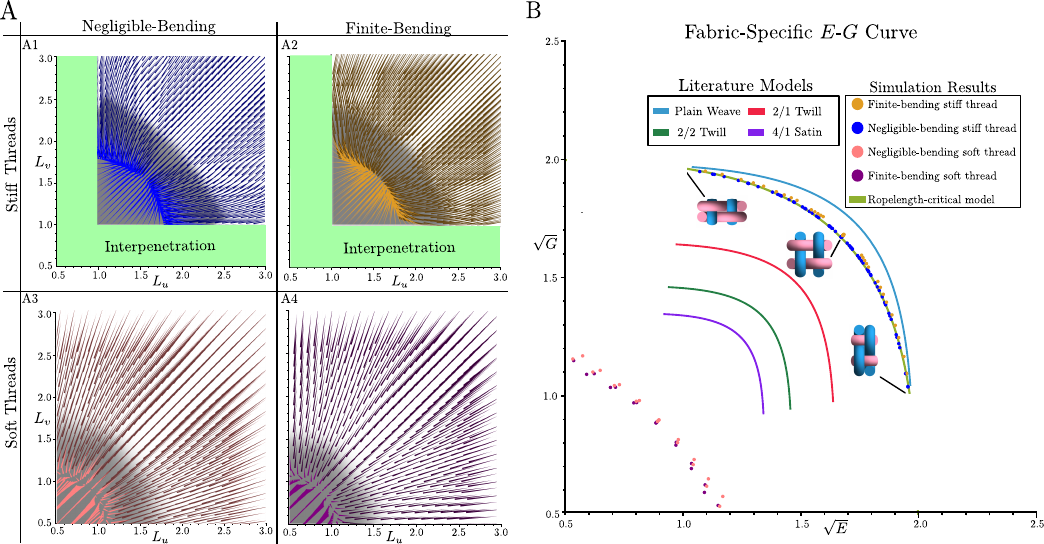}
	\caption{(A) Energy-minimizing configurations of a plain-weave unit cell for various prescribed thread rest lengths are illustrated across four different physical thread models: (A1) negligible bending modulus with stiff threads (blue), (A2) finite bending modulus of an elastic rod with stiff threads (orange), (A3) negligible bending modulus with soft threads (pink), and (A4) finite bending modulus of an elastic rod with soft threads (purple). Each tapered line represents a single energy-minimizing unit-cell configuration. The narrow end of the line points at the prescribed target thread arclengths \((L_u, L_v)\), and its wide end at the tangentially projected equilibrium unit-cell dimensions \((\sqrt{E}, \sqrt{G})\). The lengths of the stiff thread model are measured in units of thread diameter, whereas the lengths of the soft threads are not normalized because they can contract. The green area indicates the unit cell's lateral length corresponding to the interpenetration of threads. 
    Panel (B) displays the tight-weave \(E-G\) curves obtained for all physical models in (A1-A4); the \(E-G\) curve of the ropelength-critical weave obtained through numerical constrained optimization (which is expected and indeed lies closely to A1); and, literature predictions of the unit cell dimensions for different weaves (plain, twill, and satin) \cite{Peirce01031937,Love}. Insets show unit cell jammed configurations for different values of $\alpha$ of the plain weave. Although they may appear unjammed, the gap is crucial for allowing the out-of-plane thread deformation required by the weaving topology.
    }
	\label{fig: EG}
\end{figure}

The same jamming transition also imposes a nontrivial relationship between $E$ and $G$ in the tight limit, arising from the weaving topology. Intuitively, adjacent warp threads cannot be too close to each other as weft threads must be allowed to pass between them, and vice versa, which imposes a relation between the width and length dimensions of each unit cell. Such a relation was pointed out in a seminal paper by Peirce \cite{Peirce01031937}, suggesting an approximate theoretical model for tight plain weaves. This model was later generalized in \cite{Love} to account for the lateral compression of threads across various weave types, such as twills or sateens. The theoretical framework, which aligns with findings from industrial loom experiments \cite{Dickson}, demonstrated that the geometric constraints depend on the weave topology, the physical properties of the threads, and their interaction, see Fig. \ref{fig: EG}. Recent work identifies jammed states in knitted fabrics with a minimization of area per unit cell \cite{Pulling}, suggesting that these fabrics also admit an effective coupling between $E$ and $G$. However, $E-G$ relations in knits are less straightforward to model directly with an equation due to the intricate knitted topology.

The fabric-dependent jammed-state relation between $E$ and $G$ suggests that its metric (Eq. \ref{eq: def-metric}) collapses to a one-dimensional curve. One can then write the metric in the form
\begin{equation}
	\dd s^2 = E\br{\alpha\br{u,v}}\dd u^2  + G\br{\alpha\br{u,v}}\dd v^2,
	\label{eq: general-metric}
\end{equation}
where we now consider $E$ and $G$ to be known functions that are determined by the threads' physical attributes. The scalar function $\alpha(u,v)$ may vary between different locations in the fabric. It determines the position on the universal $E-G$ curve of the local jammed state at that location.

In a smart fabric, threads change (typically reduce) their rest length upon actuation. The degree of local compression is assumed to be a controllable degree of freedom in the manufacturing process, e.g., the cross-linking density at each point along a liquid crystal elastomer thread. In a wide regime of fabric parameters (discussed in the following), this actuation brings the smart fabric into the tight-weaving regime by inducing jamming and thus rigidifying the local fabric geometry as discussed above. 

To illustrate this, we simulate the mechanical response of a plain woven fabric using a three-dimensional finite element model. The model focuses on a single unit cell of the weave (see Fig. \ref{fig: EG}B), with periodic boundary conditions imposed. Each thread is represented as a chain of springs arranged according to the interlacing topology of a plain woven fabric. The elastic energy consists of a stretching term that penalizes deviation from the springs' rest lengths, and a bending term that penalizes angles between neighboring springs. The ratio between the bending and stretching moduli determines an ``effective thickness'', which depends on the physical nature of the threads and their elastic anisotropy. Additionally, the threads possess a finite physical thickness, which prevents them from coming too close to each other beyond a certain minimum distance. For a stiff thread, this minimum distance is determined by the thread’s diameter (uncontractible threads). In contrast, a soft thread can contract down to a core diameter, as described in \cite{Pulling}. Therefore, we define two sets of threads: one set consists of stiff threads, which feature a hard-cylinder repulsive term in the elastic energy that strongly penalizes inter-thread distances smaller than 1 (establishing the units for all other length scales in this model). The other set includes soft threads that exhibit compressive potential derived from a physical model described in \cite{Pulling} (with a core diameter of $0.5$); for more details, see App. \ref{app: sinulations}.

We simulate the two ``effective thickness'' models. The first model includes bending, representing threads made from an isotropic elastic material, such that the effective thickness is identical to the physical thickness. The second model includes negligible bending (still needed for regularization), representing threads made of bundled stiff fibers \cite{BarsiF, AdnanMohammed2018BboC} (effective thickness $\ll$ physical thickness). In the vanishing-bending limit with stiff threads, we expect this model to produce the \emph{ropelength-critical weave} (following \cite{CantarellaJason2006CftG}), namely the woven structure of least arclength constrained to an inter-thread distance of $1$. We conjecture that the ropelength-critical weave admits an explicit analytical formula in terms of piecewise planar-circular arcs and critical clasps, a topic for forthcoming work \cite{Berin2026inprep}. Here we do not prove this; instead, we obtain it numerically via constrained optimization (App.~\ref{appendix: Clasp}).

For each physical model, we minimize the total energy of the unit cell for a given set of rest lengths of the threads to determine the optimal rest configuration of the fabric. In particular, we retrieve the projected lengths of the threads, which translate to $2\sqrt{E}$ and $2\sqrt{G}$ in our chosen coordinate system. The results corresponding to simulations at different rest lengths of the threads are plotted in Fig. \ref{fig: EG}.

As Fig. \ref{fig: EG} shows, the tight weaving regime comprises a large domain of thread rest-lengths that, when contracted, reach the jamming transition. The metric components of such fabrics indeed collapse to a one-dimensional curve, namely a strict equality relating $E$ and $G$, which is in alignment with the metric form of Eq. \ref{eq: general-metric}.

We conclude that for tightly woven smart fabrics in the jammed state, $E(\alpha)$ and $G(\alpha)$ are known functions that are determined by the threads' physical attributes. The scalar function $\alpha(u,v)$ depends on the local rest lengths of the threads at each unit cell. 
We thus reinterpret $\alpha(u,v)$ as a local structure field that can be imposed during the manufacturing process via control over the local thread properties. Experimentally, this can be achieved either by accurately weaving with a varying-actuation-profile thread \cite{TellesLewis}, or by using two-step crosslinking \cite{Barnes2019} with the second crosslinking done post-weaving. Neither approach has been implemented yet, to the best of our knowledge. We hope that this manuscript provides the motivation for developing such a system.

Having established the relation described in Eq. \ref{eq: general-metric}, with known (fabric-dependent) functions $E(\alpha)$ and $G(\alpha)$ and assumed control over the structure field $\alpha(u,v)$, we turn to investigate the geometry of the actuated jammed fabric. Gauss's Theorema Egregium connects the metric tensor and the Gaussian curvature of a surface $K\br{u,v}$ \cite{doCarmoManfredoP2016DGoC}. The Gaussian curvature associated with Eq. \ref{eq: general-metric} is 
\begin{equation}
	K\br{u,v} =  \frac{ \br{EG}' \left(\alpha_v^2 E'+\alpha _u^2 G'\right)
		-2GE \left(\alpha_v^2 E''+\alpha _u^2 G''+\alpha_{uu} G'+\alpha_{vv}E'\right)}{4 E^2 G^2}.
	\label{eq: K of alpha}
\end{equation}
This expression readily describes the geometry of an actuated, tightly woven fabric in terms of its local structure field $\alpha(u,v)$. It thus allows predicting the resulting surface shape, solving the forward design problem. Namely, from the thread actuation profile, manifested in $\alpha(u,v)$ via a calibration curve of the type of Fig. \ref{fig: EG}, we calculate the (rigid) intrinsic geometry of the actuated fabric.

\begin{figure}[h]
	\includegraphics[width=1.0\linewidth]{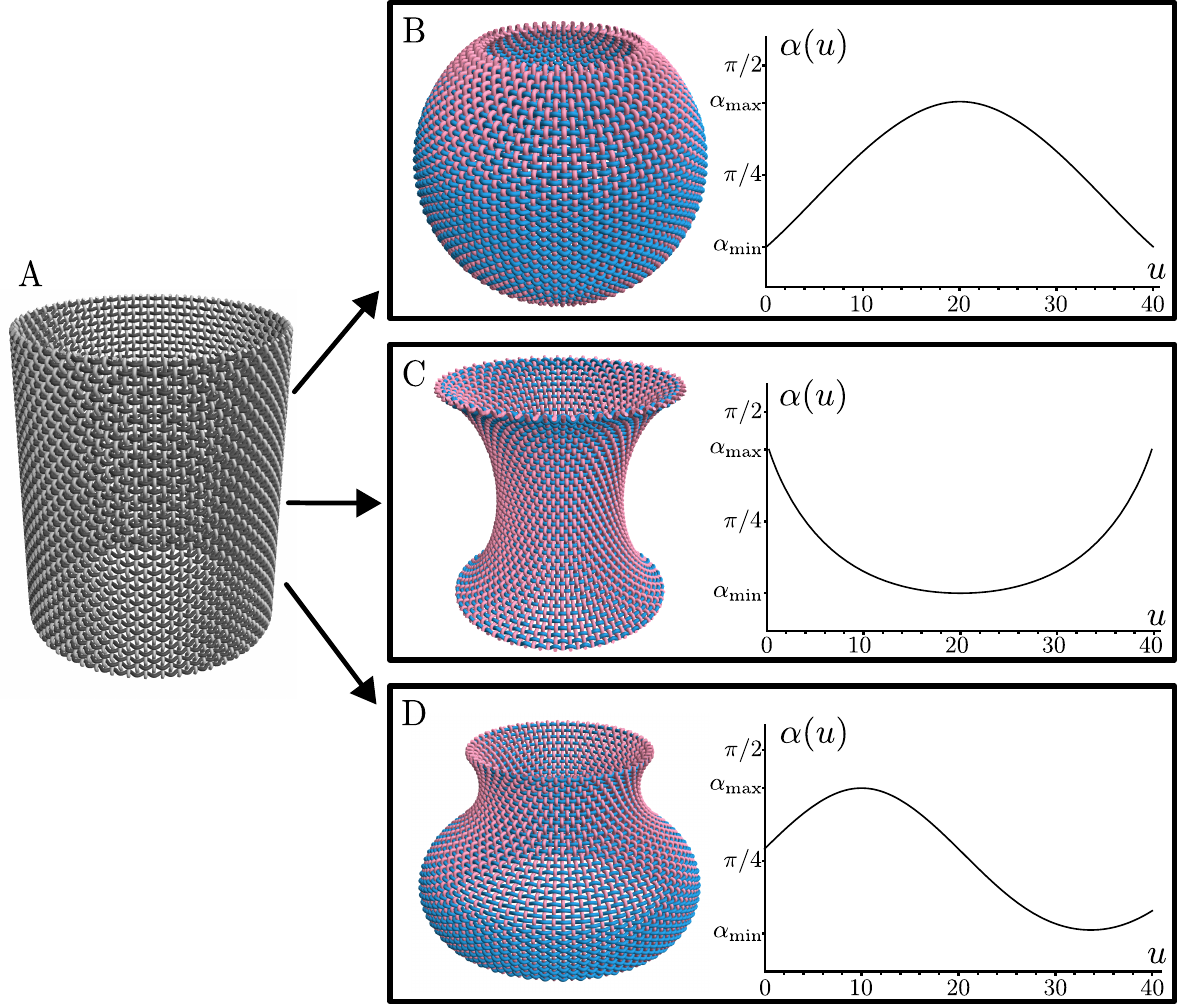}
	\caption{Inverse design of woven smart-fabric surfaces of revolution. (A) A pre-actuated woven cylinder.
		(B-D) Actuating the smart fabric with different actuation profiles $\alpha(u)$ deforms the cylinder into (B) a sphere, (C) a pseudo-sphere, and (D) a vase-shaped surface of revolution. The actuation profile required to obtain each desired shape was solved using Eq. \ref{eq: alpha-ODE}.}
	\label{fig: InverseAnalytic}
\end{figure}

We turn to study the inverse design problem. Eq. \ref{eq: K of alpha} can be rewritten as a second-order PDE for $\alpha\br{u,v}$

\begin{equation}
	G'\alpha_{uu} + E'\alpha_{vv} = \frac{\br{EG}'}{2EG} \left(\alpha_v^2 E'+\alpha _u^2 G'\right) - \alpha_v^2 E''-\alpha _u^2 G'' - {2EG}K\br{u,v}.
	\label{eq: alpha of K}
\end{equation}
In some special cases, Eq. \ref{eq: alpha of K} is analytically solvable, for example, the case of $E=\cos^2\alpha$, $G=\sin^2\alpha$ and $K\br{u,v}=\textrm{const}$ in which Eq. \ref{eq: alpha of K} becomes the Sine-Gordon equation. Such solutions are not likely to be useful since $E(\alpha)$ and $G(\alpha)$ are typically calibrated experimentally, and $K(u,v)$ represents an arbitrary target surface.
In some cases, additional symmetries allow for significantly simplifying   Eq. \ref{eq: alpha of K}. This is the case for target surfaces of revolution, where we assume that the thread structure shares the same symmetry, namely that threads lie along latitudes and longitudes of the surface. The inverse design problem reduces to a second-order ODE,
\begin{equation}
	\alpha''(u)
	+A\left[\alpha\right]\,\alpha'(u)^2
	=B\left[\alpha\right]\,K\left(\int\sqrt{E(\alpha(u)))}\, du\right),
	\label{eq: alpha-ODE}
\end{equation}
where $A\left[\alpha\right]=\frac{d}{d\alpha}\log\left(\frac{G'}{\sqrt{E G}}\right)$ and $B\left[\alpha\right]=-\frac{2E G}{G'}$.
Given initial conditions for $\alpha(u)$, Eq. \ref{eq: alpha-ODE} can be solved, either analytically ( App. \ref{appendix: ForwardSymmetric}) or numerically. In Fig. \ref{fig: InverseAnalytic} we solve Eq. \ref{eq: alpha-ODE} for the model of negligible bending with stiff threads ( App. \ref{app: fits}) on several surfaces of revolution. We will use this model as a showcase for the rest of this work.

Nonetheless, the general case makes a well-posed hyperbolic problem. For common fabrics such as woven fabrics \cite{StructureandMechanicsofWovenFabrics} and knitted fabrics \cite{Pulling}, the effect of the volume excluded by each thread on the trajectories of crossing threads implies that $dE/dG<0$.
Without loss of generality, we choose $E(\alpha)$ to be a monotonically decreasing function, and $G(\alpha)$ to be a monotonically increasing function. Thus, Eq. \ref{eq: alpha of K} is a hyperbolic PDE on $\alpha(u,v)$ with characteristic curves that are transverse to both families of threads. Of course, the PDE depends on the value of the target Gaussian curvature at each point. To complete a well-posed Cauchy initial-value problem, the hyperbolic PDE requires a Cauchy-type initial condition \cite{PinchoverYehuda2005AItP}. Thus, the problem is fully set given the following:
(i) a target surface, (ii) a curve on the target surface, and (iii) a shrinkage profile along the curve. The chosen curve is destined to be the actuated shape of one of the threads, $v=0$. The profile $\alpha(u,0)$ along the curve and the curve's geodesic curvature completely determine the perpendicular derivative $\partial_v\alpha(u,0)$ ( App. \ref{appendix: PseudoCode}), and thus complete the setting of a Cauchy problem. This guarantees a unique solution for $\alpha(u,v)$ in the surroundings of the initial curve, which may extend all the way to the boundaries or diverge at some finite horizon \cite{PinchoverYehuda2005AItP}.

The solution to a Cauchy problem, as stated above, is a solution to the inverse design problem of smart fabrics. Given a target surface, we obtain an explicit function $\alpha(u,v)$, which can then be transcribed to the explicit shrinkage of each thread at each point using Eq. \ref{eq: alpha of K}. Therefore, given a robust and precise technique for the manufacturing of smart fabrics with given thread-shrinkage profiles, we provide an explicit recipe for fabrics that will deform into an arbitrary target geometry upon actuation, at least locally.

For practical implementations of the inverse problem, the target surface will be given in a discrete form, such as a triangular mesh. An algorithm for numerically solving the initial value problem described above on such a mesh is presented in App. \ref{appendix: PseudoCode}, in the form of pseudo-code. Such a solution is not guaranteed to extend to the full mesh. An alternative approach is to search for a global coordinate parameterization of the target surface, in which the metric of the surface at any point takes the form   Eq. \ref{eq: general-metric} for given functions $E$ and $G$. Since a global solution is not guaranteed, the idea is to find a parametrization that best approximates the local rules. Finding surface parametrizations that approximate different local relations is an active field of study in computer graphics and geometric processing. Many methods have been developed to address similar problems, such as the local/global approach \cite{LiuLigang2008ALAt}, the commuting vector fields approach \cite{MirelaVaxman}, and others.

\begin{figure}[h]
	\includegraphics[width=1.0\linewidth]{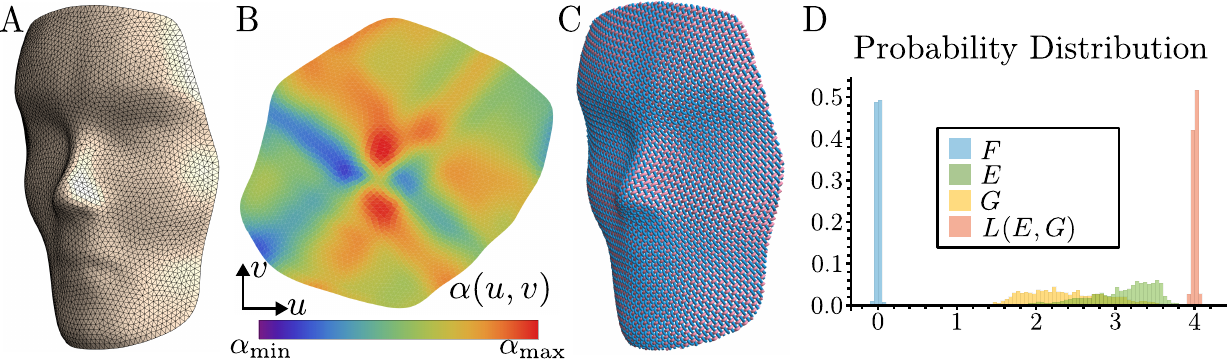}
	\caption{Inverse design of a generic woven smart-fabric surface. {(A)} Target face geometry given as a triangular mesh.
		{(B)} An approximate solution of  Eq. \ref{eq: wovenfabric-metric} obtained numerically using the adapted Corman-Crane algorithm (App. \ref{appendix: Matlab}). The algorithm outputs both the weaving domain and the actuation parameter $\alpha(u,v)$ at each point, needed for manufacturing the smart fabric.
		{(C)} Upon actuation, the initially planar fabric will adopt the geometry of the target face shape, with the shown thread pattern.
		{(D)} Histogram of metric elements confirming the numerical solution. The sharp distribution of the $F$ component at $0$ confirms orthogonality. The function $L(E,G)$ is a specific norm function on the target metric components, such that the value $L=4$ confirms adherence to the imposed tight-weave $E-G$ curve ( App. \ref{app: fits}).}
	\label{fig: resultface}
\end{figure}

Recently, a numerical approach that is very natural to our problem was presented by Corman and Crane  \cite{Corman:2025:RSPCode}. The approach is designed to find rectangular surface parameterizations, namely those with $F=0$. The method establishes integrability conditions for an orthogonal local frame field to represent a coordinate-derived frame. Their algorithm then searches for an optimal parameterization with respect to these conditions and any additional energy term written in terms of the local frame. For our purposes, we simply construct an energy penalty for deviating from the target metric of the model of negligible bending with stiff threads. This requires an adaptation of the code used in \cite{Corman:2025:RSPCode} ( App. \ref{appendix: Matlab}). Approximate solutions to the inverse design problem are shown in Fig. \ref{fig: resultface}.

In this example, the target surface is a face represented as a triangular mesh. 
The Corman–Crane method produces a rectangular parameterization consistent with 
the tight-weave constraints, to which threads can be aligned (Fig. \ref{fig: resultface}). 
The actuation parameter $\alpha$ varies across the surface, evident from the differing 
eccentricity of the rectangular patches in the chin region compared to the nose. 
This variation is also visualized in the parametrization domain, where each face is 
colored by its $\alpha$ value (Fig. \ref{fig: resultface}{B}). The histogram of the
metric components on each face, (Fig. \ref{fig: resultface}{D}), confirms that the parametrization is 
orthogonal and satisfies the constraint $L(E,G)=4$ ( App. \ref{app: fits}) confirming with Eq. \ref{eq: general-metric} of the model negligible bending with stiff threads. Together, these results 
demonstrate that the discrete construction provides a faithful, though approximate, 
realization of the target geometry.

\section*{Discussion}
In this work, we establish a geometric framework for the inverse design of tightly woven smart fabrics. In the tight-weaving regime, the mechanical jamming of threads imposes an orthogonal coordinate system with strictly coupled metric components. Through finite element simulations of weaving unit cells, we validate that this coupling follows a universal master curve. For a range of physical models, this curve can be very well approximated by an analytic curve ( App. \ref{app: fits}). Based on this constitutive relation, we formulate the inverse design problem as a hyperbolic partial differential equation for the local actuation field $\alpha(u,v)$. We provide constructive solutions for this system, ranging from an exact analytical solution for surfaces of revolution to a global optimization method for general discrete surfaces.

Practical realizations of active woven textiles have recently been demonstrated by Silva et al., who successfully integrated liquid crystal elastomer (LCE) filaments into standard weaving patterns \cite{Terentjev}. Their approach used a hybrid structure in which one set of threads (the weft) is active and exhibits a uniform shrinkage profile upon heating, while the orthogonal set (the warp) remains passive. A particularly relevant case is their polar weave design, which is fabricated as a flat disk and transforms into a cone, effectively forming a positive curvature singularity at the tip. While the threads remain orthogonal, the radial divergence of the passive warp threads implies that the inter-thread spacing increases linearly with the distance from the tip. Consequently, such a structure cannot comply with the tight-weaving condition described in our model on an extended domain. A polar weaving system could still be described by our model (App. \ref{appendix: General Gaussian Curvature}) if radial threads are constantly added during weaving away from the center to maintain the tight structure.

Although we considered the simple case of plain woven fabrics, our assumptions also hold for more complex fabrics in the tight regime. The majority of textile structures possess a symmetry that excludes spontaneous shear, and the jamming transition is expected to eliminate an additional degree of freedom. The calibration curve Fig. \ref{fig: EG} for such fabrics may be more complicated and harder to model. Nonetheless, once it is achieved (if only by experimental measurement), our suggested inverse design protocols may be readily implemented.

Nonetheless, implementing our model for knitted and other fabrics may require addressing an additional component. The weaving structure preserves coarse-grained symmetry with respect to normal reflections and thus prefers no extrinsic curvature; Conversely, knitted fabrics exhibit a preferred non-zero saddle-like extrinsic curvature that stems from the knitting structure \cite{NiuLauren2025Gmok}, and is responsible for the fabric's tendency to strongly curl at its boundaries \cite{kurbak2008basic}. Utilizing this phenomenon grants the manufacturer access not only to the fabric's intrinsic geometry but also, to some degree, to its extrinsic curvatures. This, in turn, may allow control over the resulting shape choice among all isometries of the target surface \cite{Aharoni7206}.

Contrary to local solutions to the inverse problem, global solutions do not always exist. Nonetheless, as shown in several similar problems \cite{MirelaVaxman}, global solutions are expected to exist if they are allowed to contain point disclinations, around which the warp and weft threads switch roles, creating a 90-degree surplus/deficit angle. We are not aware of a weaving mechanism that allows for introducing such objects into a woven fabric; however, if such a mechanism exists, global inverse design is readily feasible with our numerical inverse design scheme.

\section*{Data Availability}
All code used to generate data that support the findings of this study is openly available in Zenodo (see \cite{einav143_2026_18324284}) and in the supplementary information of this article.

\begin{acknowledgments}
We thank Etienne Corman and Keenan Crane for clarifications and advice on adapting their code. We thank Myfanwy E. Evans and Randall D. Kamien for fruitful discussions. This research was supported by the Israel Science Foundation (grant No. 2677/20).
\end{acknowledgments}

\appendix
\counterwithin{figure}{section}

\section{Numeric Scheme}
\label{app: sinulations}
To validate this geometric limit and explore the behavior of physical threads with finite thickness and bending stiffness, we implemented a custom numerical simulation (code available at \url{https://doi.org/10.5281/zenodo.21739296}). The simulation operates on a minimal periodic unit cell of the fabric, consisting of interwoven orthogonal row and column threads with fixed target lengths $L_u, L_v$. Each thread is discretized into a sequence of nodes connected by straight segments. The initial configurations of the threads are prescribed as interlaced sinusoidal curves to mimic the alternating topology of a plain weave. Periodic boundary conditions are enforced by requiring the endpoints of threads to lie on the lateral plane (the $z=0$ plane).
The structure is governed by an elastica energy formulation, which accounts for stretching, bending, and inter-thread steric repulsion. The total energy is the sum of these contributions: $E_{\text{total}} = E_{\text{stretch}} + E_{\text{bending}} + E_{\text{repulsion}}$. To ensure that threads closely follow their prescribed target lengths, we apply a very large stretching stiffness, heavily penalizing length deviations via $E_{\text{stretch}} = \frac{1}{2} k_s \sum_i (l_i - l_{i0})^2$, where $\sum_i l^{u/v}_{i0}=L_{u/v}$. The bending energy penalizes high curvatures and is defined as $E_{\text{bending}} = \frac{1}{2} k_b (2r)^2 \sum_j (1 - \cos\theta_j)$, where $\theta_j$ is the angle between consecutive segments, $k_b$ is the bending stiffness, and $r$ is the thread radius. The total energy is minimized using the L-BFGS numerical optimization algorithm.
In our studies, we model two sets of threads to demonstrate how the physical elastic fabric converges to the purely geometric, jammed ropelength-critical state derived above. The first set consists of soft threads, for which we employ the soft-shell, hard-core repulsion potential detailed in \cite{Pulling}. Adopting their parameters, this compression energy is modeled as an inverse power law and takes the discrete form $E_{\text{repulsion}} = k \frac{(2r - 2r_{\text{core}})^2}{p(p-1)} \sum \left[ \zeta^{1-p} - 1 - (p-1)(1-\zeta) \right]$ for node pairs where the distance $d < 2r$. Here, $\zeta = (d - 2r_{\text{core}})/(2r - 2r_{\text{core}})$, with $r_{\text{core}}$ being the radius of the incompressible inner core, and $p$ is the power-law exponent. The second set consists of stiff threads, where we instead use a stiff constraint potential to heavily penalize the squeezing of the thread. This repulsion takes the form $E_{\text{repulsion}} = \varepsilon \sum \log(1 + \exp((2r - d)/0.01))^2$, providing a robust penalty against interpenetration.

\section{The Ropelength Problem} 
\label{appendix: Clasp}

Ropelength problems seek the configuration of a knot or link that minimizes its total arc length while maintaining a unit-thickness constraint \cite{CantarellaJason2006CftG, sullivan2005some}. Guided by this picture, we identify the tightly woven fabric unit cell as a critical ropelength problem, with the weaving topology and periodic boundary conditions in mind. We focus on a unit cell of woven fabric, in which the periodic boundary conditions are imposed by requiring that all endpoints lie on the same plane. We further fix the aspect ratio between the end-to-end distances of the $u$-threads and $v$-threads. We then use numerical constrained optimization with a hard minimal-distance constraint between the two curves. We employ a \emph{Sequential Least Squares Programming (SLSQP)} algorithm, which iteratively approximates the arclength objective function and the distance constraints as a quadratic program, allowing the system to converge to the jamming transition where the threads are in a state of minimal length subject to the constraint. 

We find that the length-optimal curves decompose into regular circular arcs and segments of generalized clasp solutions \cite{ sullivan2005some}. This is analogous to the ropelength-critical solution for the Borromean rings, which similarly comprises piecewise analytic segments rather than a single smooth curve, as discussed in \cite{CantarellaJason2006CftG}. The full ropelength-oriented analysis of fabrics is the topic of a separate, forthcoming work \cite{Berin2026inprep}.

\section{Analytical Approximation of the E-G Master Curve}
\label{app: fits}
\begin{figure}[!ht]
    \centering
    \includegraphics[width=0.5\linewidth]{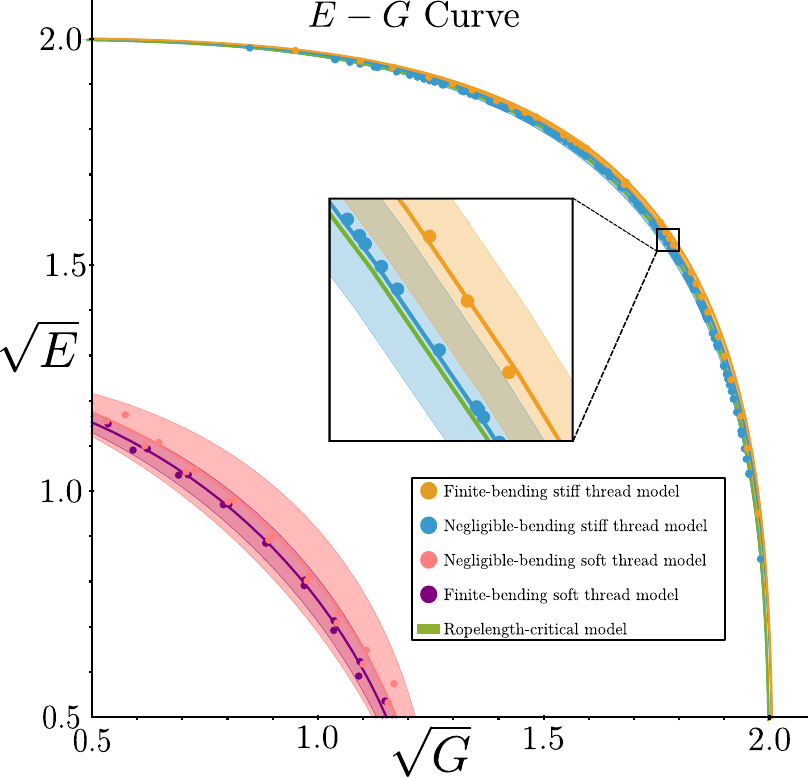}
    \caption{The tight-weave $E-G$ curves obtained for all physical models, fitted to the one-parameter form Eq. \ref{eq: wovenfabric-metric} (Table. \ref{tab: cfit} for the $c$ value and the $R^{2}$ values). Also shown is the E-G curve of the ropelength-critical weave, obtained numerically by constrained optimization. As expected, it lies close to the negligible-bending, stiff threads' curve.}
    \label{fig: fit EG}
\end{figure}

A relatively simple relation that makes a good approximation for the models we've simulated, and many other physical thread structures, reads ( Fig. \ref{fig: fit EG})
\begin{equation}
	\sqrt{E}=D\cos^c\alpha(u,v)\, ,\quad
	\sqrt{G}=D\sin^c\alpha(u,v)\, ,\quad
	\textrm{where }\,\alpha(u,v)\in \br{0,\frac{\pi}{2}}.
	\label{eq: wovenfabric-metric}
\end{equation}
Here $D$ is the sum of the warp and the weft diameters, and $c$ is a fitting parameter.
If one assumes that the threads are laterally incompressible (stiff threads), then $\alpha$ is further restricted such that $E,G \ge 1$. For soft threads, $E,G \ge D^2_{\text{core}}$ for minimal contraction diameter. In each model, we fit $c$ ( Table \ref{tab: cfit}). We find good agreement at about $c=0.5$ for the stiff-threads models.
\\
\begin{table}[!ht]
    \centering
    \caption{Fitted $c$ parameter for Eq. \ref{eq: wovenfabric-metric} for the simulation models}
    \label{tab: cfit}

    \begin{tabular}{|c|c|c|}
        \hline
        & Negligible Bending & Finite Bending \\ \hline
        Stiff Threads & $c=0.50\pm0.01,\ R^2=99.5\%$ & $c=0.52\pm0.01,\ R^2=99.86\%$ \\ \hline
        Soft Threads  & $c=1.00\pm0.17,\ R^2=98.1\%$ & $c=0.97\pm0.40,\ R^2=84.4\%$ \\ \hline
    \end{tabular}
\end{table}

\section {Gaussian Curvature Formulae}
\label{appendix: General Gaussian Curvature}
Gauss’s \emph{Theorema Egregium} states that the Gaussian curvature $K$ of a two-dimensional surface is an intrinsic invariant that can be computed directly from the metric tensor and its derivatives.

Given a general two-dimensional  metric 
\[
\dd s^2 = E(u,v)\,\dd u^2 + 2F(u,v)\,\dd u\,\dd v + G(u,v)\,\dd v^2,
\]
the Gaussian curvature is given by the Brioschi formula:
\[
K = \frac{-1}{2\sqrt{EG - F^2}} \left[ \frac{\partial}{\partial u} \left( \frac{G_u - 2F_v + E_v}{\sqrt{EG - F^2}} \right) + \frac{\partial}{\partial v} \left( \frac{E_v - 2F_u + G_u}{\sqrt{EG - F^2}} \right) \right].
\]

As we discuss in the manuscript, the general metric describing the geometry of a tight fabric is
\[
\dd s^2 = E\br{\alpha\br{u,v}}\dd u^2  + G\br{\alpha\br{u,v}}\dd v^2.
\]
The Gaussian curvature is then
\[K\br{u,v} =  \frac{ \br{E G'+E'G} \left(\alpha_v^2 E'+\alpha _u^2 G'\right)
	-2GE \left(\alpha_v^2 E''+\alpha _u^2 G''+\alpha_{uu} G'+\alpha_{vv}E'\right)}{4 E^2 G^2}.
\]
The metric tensor that describes polar weaving is, in polar coordinates:
\[
\dd s^2 = E\br{\alpha \br{r,\theta}}\,\dd r^2  + r^2 G\br{\alpha \br{r,\theta}}\,\dd \theta^2.
\]
The Gaussian curvature is then
\begin{align*}
	K\br{r,\theta} & =  \frac{ \br{E G'+E'G} \left(\alpha_\theta^2 E'+r^2\alpha _r^2 G'\right)
		-2GE \left(\alpha_\theta^2 E''+\alpha _r^2 r^2 G''+\alpha_{rr} r^2 G'+\alpha_{\theta\theta}E' +2\alpha_r r G'\right)}{4 r^2 E^2 G^2 }\\
	& +\frac{\alpha_r E' \br{2G+rG'}}{4rG E^2}.
\end{align*}

\section{Geometric construction of a local solution for the Cauchy problem}
\label{appendix: PseudoCode}
The inverse design problem for a tight smart fabric whose local deformation obeys the metric
\begin{equation}
	\dd s^2 = E\br{\alpha\br{u,v}}\dd u^2  + G\br{\alpha\br{u,v}}\dd v^2,
	\label{eq: general-metric A}
\end{equation}
can be written as a hyperbolic PDE
\begin{equation}
	G'\alpha_{uu} + E'\alpha_{vv} = \frac{\br{EG}'}{2EG} \left(\alpha_v^2 E'+\alpha _u^2 G'\right) - \alpha_v^2 E''-\alpha _u^2 G'' - {2EG}K\br{u,v}.
	\label{eq: alpha of K A}
\end{equation}
The associated Cauchy problem requires, in addition to Eq. \ref{eq: alpha of K A}, also the value of $\alpha\br{u,0}$ and $\alpha_v\br{u,0}$ on an initial curve $v=0$ (transverse to the characteristic curves). $\alpha_v\br{u,0}$ can be calculated from $\alpha\br{u,0}$ and the geodesic curvature of the initial curve   $\kappa_{g}(u)$ via the relation \cite{doCarmoManfredoP2016DGoC}
\begin{equation*}
	\alpha_v\br{u,0} = -\frac{2\kappa_{g}(u)E\br{\alpha(u,0)}\sqrt{G\br{\alpha(u,0)}}}{E'\br{\alpha(u,0)}}.
	\label{eq: geodesic curvature an av}
\end{equation*}
Geometrically, the solution is constructed as follows: we start with an arbitrary curve on the target surface, which we identify with the $u$-curve $v=0$. We arbitrarily choose $\alpha(u,0)$, which in turn defines scale marks along the initial curve, separated by the distance $\sqrt{E\bigl(\alpha(u,0)\bigr)}$. From each mark, we emit a $v$-curve, perpendicular to the initial curve, at speed $\sqrt{G\bigl(\alpha(u,0)\bigr)}$. 
The loci along these $v$-curves after an infinitesimal time $\epsilon$ make the $u$-curve $v=\epsilon$. Iterating this construction produces a local rectangular coordinate system on the surface, which obeys the metric  Eq. \ref{eq: general-metric A}. The procedure breaks down when two such $u$-curves intersect (namely $E(\alpha)=0$) or when two $v$-curves intersect ($G(\alpha)=0$). The geometric construction is summarized in Algorithm \ref{alg: PDElocalSol}.

\begin{algorithm}[H]
	\caption{Geometric Construction of a Local Solution to the Cauchy Problem}
	\label{alg: PDElocalSol}
	\begin{algorithmic}[1]
		\State \textbf{Input:} 
		\begin{itemize}
			\item Target surface,
			\item Initial curve $\gamma_0$ on the surface,
			\item Initial profile $\alpha(u,0)$ along $\gamma_0$.
		\end{itemize}
		\State \textbf{Output:} Surface parameterization $(u,v)$ and structure field $\alpha\br{u,v}$, in which the metric of the target surface  Eq. \ref{eq: general-metric A}.
		\State Parametrize the curve $\gamma_{0}(u)$ such that 
		\(
		\gamma_{0}'(u) = \sqrt{E\bigl(\alpha(u,0)\bigr)}.
		\)
		\State Set $v \gets 0$.
		\State Set positive $\delta v \ll$ surface dimensions and curvature scales.
		\While{$\forall u$:~$E\br{\alpha(u,v)} > 0$ and $G\br{\alpha(u,v)} > 0$}
		\For{all u}
		\State $\gamma_{v+\delta v}(u) = \exp_{\gamma_v(u)}\left(\sqrt{G\bigl(\alpha(u,v)\bigr)}\delta v\cdot\widehat{\gamma_v'(u)^\perp}\right)$ \qquad (where $\exp$ is the exponential map)
		\EndFor
		\State Set
		\(
		\alpha(u,v+\delta v) = E^{-1}\bigl(\gamma_{v+\delta v}'(u)^2\bigr).
		\)
		\State Set $v \gets v+\delta v$.
		\EndWhile
	\end{algorithmic}
\end{algorithm}

\section{Inverse Problem Symmetric Solution}
\label{appendix: ForwardSymmetric}
In the symmetric case of a surface of revolution, Eq.~\ref{eq: alpha-ODE} reads
\begin{equation}
	\alpha''(u) + A[\alpha(u)]\,\alpha'(u)^{2}
	= B[\alpha(u)]\,K\bigl(s(u)\bigr),
	\qquad
	s(u) = \int_{0}^{u}\sqrt{E\bigl(\alpha(\tilde u)\bigr)}\,\dd\tilde u,
	\label{eq:app-6}
\end{equation}
with
\[
A[\alpha] = \frac{\dd}{\dd\alpha}\log\!\left(
\frac{G'(\alpha)}{\sqrt{E(\alpha)G(\alpha)}}
\right),
\qquad
B[\alpha] = -\,\frac{2E(\alpha)G(\alpha)}{G'(\alpha)}.
\]

Define
$$
	w(u) \coloneqq
	\alpha'(u)\,
	\frac{G'\bigl(\alpha(u)\bigr)}{\sqrt{E\bigl(\alpha(u)\bigr)G\bigl(\alpha(u)\bigr)}}.
$$
Then Eq.~\ref{eq:app-6} is equivalent to the first-order system
\begin{equation}
	\begin{aligned}
		\alpha'(u)
		&= w(u)\,
		\frac{\sqrt{E\bigl(\alpha(u)\bigr)G\bigl(\alpha(u)\bigr)}}
		{G'\bigl(\alpha(u)\bigr)},\\[4pt]
		s'(u)
		&= \sqrt{E\bigl(\alpha(u)\bigr)},\\[4pt]
		w'(u)
		&= -\,2\sqrt{E\bigl(\alpha(u)\bigr)G\bigl(\alpha(u)\bigr)}\,
		K\bigl(s(u)\bigr),
	\end{aligned}
	\label{eq:app-first-order-system}
\end{equation}
with initial data
\[
\alpha(0) = \alpha_{0},\qquad
s(0) = s_{0},\qquad
w(0) = w_{0}
= \alpha'(0)\,
\frac{G'(\alpha_{0})}{\sqrt{E(\alpha_{0})G(\alpha_{0})}}.
\]

For the special case of constant Gaussian curvature \(K(s)\equiv K_{0}\), the system Eq.
\ref{eq:app-first-order-system} admits a first integral
\begin{equation}
	w(\alpha)^{2} + 4K_{0}\,G(\alpha) = C,
	\label{eq:app-first-integral}
\end{equation}
so that \(\alpha(u)\) is given implicitly by the integral
$$
	u - u_{0}
	= \pm \int_{\alpha_{0}}^{\alpha(u)}
	\frac{G'(\tilde\alpha)}
	{\sqrt{E(\tilde\alpha)G(\tilde\alpha)
			\bigl(C - 4K_{0}G(\tilde\alpha)\bigr)}}\,
	\dd\tilde\alpha,$$
with the constant \(C\) fixed by the initial slope \(\alpha'(0)\) via Eq. \ref{eq:app-first-integral}.

\section{Rectangular Parametrization Algorithm} 
\label{appendix: Matlab}
To generate global parameterizations on discrete triangular meshes (as shown in Fig. 4), we utilized the \textit{Rectangular Surface Parameterization} framework developed by Corman and Crane \cite{Corman:2025:RSPCode}. Their algorithm optimizes for an orthogonal coordinate system aligned with a target frame field.

We adapted their open-source implementation to enforce the specific geometric constraints of tightly woven fabrics. The original objective function minimizes standard distortion metrics (e.g., conformality or area distortion). We replaced this energy term in the file \texttt{objective\_ortho\_param.m} with a penalty that enforces the specific coupling between the metric components $E$ and $G$ derived from our physical model:

\begin{equation*}
	E^{1/c} + G^{1/c}= \text{Constant}
\end{equation*}
The modification involves calculating the error potential \texttt{err\_diag} and its associated Jacobian derivatives to guide the optimizer toward the tight-weave solution. The MATLAB implementation of this specific constraint is as follows:

\begin{lstlisting}[language=Matlab, basicstyle=\small\ttfamily, breaklines=true]
	% Enforce physical constraint: E^c + G^c = 2
	err_diag = log(exp(-2*c*ut*integral_tri - 2*c*vt*integral_tri)/2 + ...
	exp(-2*c*ut*integral_tri + 2*c*vt*integral_tri)/2);
	
	% Jacobian calculations for the optimizer
	da = -2*c*ones(Src.nf,1)/3;
	db = (2*c/3)*(exp(4*c*vt*integral_tri) - 1)./(exp(4*c*vt*integral_tri) + 1);
	daa = zeros(Src.nf,1);
	dab = zeros(Src.nf,1);
	dbb = (16*(c^2)/9)*exp(4*vt*c*integral_tri)./(exp(4*vt*c*integral_tri) + 1).^2;
\end{lstlisting}

\bibliographystyle{unsrt}
\bibliography{prl_references.bib}

\end{document}